# Experimental characterization of negative refractive index material (NRM) at Ka- band.

Sougata Chatterjee [#1], Shantanu Das [*2]

[#1] NCRA-TIFR (Tata Institute for Fundamental Research) -Pune, India.
[*2] Reactor Control Division BARC (Bhabha Atomic Research Center) Mumbai, India

*Abstract-* **In this paper, we discuss the experimental characterization of a negative refractive material (NRM) at Ka band using LR (labyrinth Ring) and wire array (WA). We describe in detail the the LR and wire array characterization separately, and after that the combined experimental results, for NRM are reported. The LR's analytical and simulation study is not new but design in Ka band and different experimental procedure for the characterization of the negative refractive index is the novelty of this paper. For performing a negative refractive index experiment we made prism of $15^0$ (Prism angle) . We get enhanced transmittance of more than 20 dB from background (reading), at a negative angle of refraction. The values of the negative refractive index in a band of about 1 G Hz (around 31 GHz) are retrieved from the experimental data.**

## I. INTRODUCTION

The left handed material (LHM) exhibits counterintuitive phenomena like a reversal of Snell's law, reversed Doppler effect and reversed Cerenkov radiation. This type of material is not found in nature where, the material permittivity and permeability ought to be simultaneously negative [1]. The LHM is called meta-material, meaning 'beyond material', which presently are not available in nature. Some typical characteristics are also discussed by Sougata et al [2-7] where the authors discuss the behavior of single photon inside a metamaterial. The possibility of realization of negative permittivity at microwave frequency was reported in [8] 1996 using thin metallic wire array, while the realization of negative permeability was subsequently reported by the same researchers J. B. Pendry et al. [9] , in 1999 using Split Ring Resonator (SRR) and Swiss Roll (SR). The experiment of negative index of refraction (NRI) was first demonstrated by D. R. Smith et al [10] in 2001 and started a new field of structure material which shows a negative refractive index (in a frequency band). Many authors are reported in different types of structures and verified the LHMs properties using different types of experiments [11-14]. The Labrynth ring (LR) is the improved version of SRR ( it has double cut compared to single cut of SRR). The first LR is proposed by Irfan Bulu et al [15] and its analytical expression is given by T. Roy et al [16]. The LR has the greater bandwidth for negative permeability, and its effective relative permeability is more comparable to any other MNG material [16].

Based on this hypothesis for LR we selected LR as our candidate for MNG material. The substrate printed wire array produced ENG (Epsilon NeGative) material, and with their combined effect (MNG and ENG) we get the DNG (Double NeGative) material where the material permeability and permittivity become, simultaneously negative in the resonant frequency zone.

In the first part we will discuss the experimental verification on both MNG, ENG, and DNG material magnitude (dB) value and phase (degree) and phase reversal information of the transmitted signal. In the next part will discuss the NRI (Negative Refractive Index) experimental of the DNG material. The retrieval method is taken from S-parameter retrieval method which is proposed by D. R. Smith et.al [17]. Thereafter, we will discuss the calculation of NRI material, followed by conclusions and references. For this experiment we have used two Horn antenna (Ka Band) and kept them at a distance of several wavelengths away from the DNG slab (LR and Wire); one antenna is transmitted and other a receiver. The receiving horn antenna is connected through Agilent's E8363B PNA Series Vector Network

Analyzer (VNA) series. For proper shielding, from background EM noise pickups, we have covered our experimental setup with microwave absorber (purchased from Orbit/FR) . The novelty of this paper is to give a new experimental arrangement and give the proper description of calculating the negative refractive index value using Prsim experiment using the geometrical expansions.

## II. LR and Metallic rod experiment

The unit cell structures of LR and wire are shown in Fig.1 (a) where the magnetic field is normal and the electric field is perpendicular to the plane of the unit cell. The analytical expression of the LR structure unit cell is [16];-

$$\mu_{reff} = 1 - \frac{F}{1 + i\frac{2R_c v}{\omega r \mu_0 (N-1)} - \frac{12v}{\pi r^2 \mu_0 \omega^2 (\pi r C_1 + 6NC_2)(N-1)^2}} \tag{1}$$

Where $N$ is the number of metallic inclusions, $\mu_0$ is free space permeability, $R_C$ is the resistance of the metal strips per unit circumferential length and $C_1$ is the distributed capacitance between the adjacent strips. The expression for $C_1$ is given in [16] is : $C_1 = (\varepsilon_0 / \pi)\ln[wN/d(N-1)]$, where $\varepsilon_0$ ispermittivityace permittivity, $w$ is the strip width and $d$ is the separation between the strips of the aorjacent rings, $r$ is the inner radius of LR , $d$ is the separation between the two rings, and $g$ is the gap between the same ring. $C_2$ is the gap capacitance between the two strips and $C_{2=}\varepsilon_0 w t_m / g$ for LR. Similarly metal printed Wire substrate can produce negative permittivity below its cutoff frequency region [8]. The schematic representation of unit cell of LR and Wire is given in Fig.1 . The Fig 1 (c) represents the combined LR and WA structure for realizing DNG material for NRM at Ka-band.

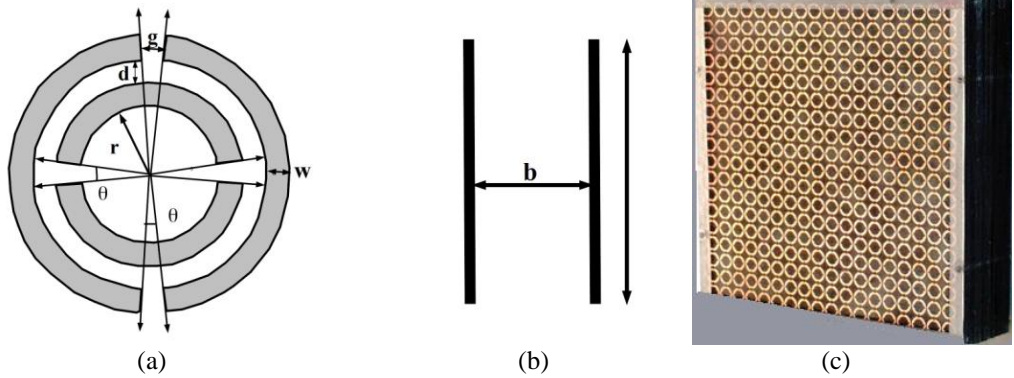

(a)  (b)  (c)

**Fig.1 Unit Cell Structure of (a) LR.(b) Wire imprinted on the substrate.(c) LR-Wire arrays.**

The fabricated structure dimensions of LR and WA are given in Table-1 and Table-2.

**Table: -1**

| | |
|---|---|
| Radius (r) | 2.1 mm |
| Width (w) | 0.3 mm |
| Gap (g) | 0.3 mm |
| Separation of two strips (d) | 0.3 mm |
| Vertical lattice distance (v) | 1.5 mm |
| Horizontal lattices constant (h) | 5.0 mm |

**Table: -2**

| | |
|---|---|
| Separation between the wire (b) | 2.0 mm |
| Wire width (w) | 0.3 mm |
| Vertical lattice distance (v) | 1.5 mm |
| Horizontal lattices constant (h) | 5.0 mm |

The structures are fabricated on Rogers RT Duriod 5880 substrates (dielectric constant 2.2 and substrate height is 0.25mm). The 20x20 LR rings are placed in two dimensional and similarly in the Wires array are formed. There are 20 such arrays that are placed in a vertical direction with the spacing of 1.5 mm given in Fig.1 (c). The experimental arrangement for characterization of the MNG (LR), ENG (WA), and DNG (LR, WA combination) is given in Fig.2 (a), (b), and (c). Fig.2 (d) gives the actual experimental arrangement of MNG, ENG and DNG material.

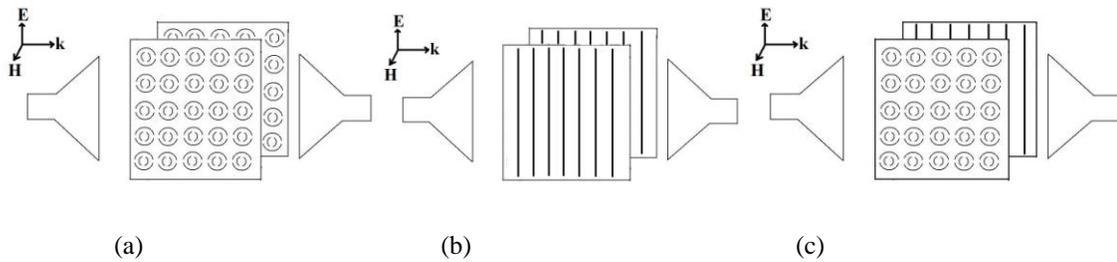

(a)　　　　　　　　　　(b)　　　　　　　　　　(c)

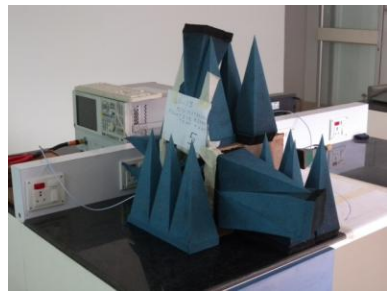

(d)

**Fig.2. Experimental arrangement (a) LR(b) Wire(c) LR and Wire and (d) Actual experiment arrangements.**

For performing the transmission experiment of LR, WA, and LR WA combination the distance between the transmission and receiving antenna is 40 cm ; in term of wave length it is $41 \times \lambda_0$ where the value of $\lambda_0 = 0.97$cm (central frequency 31GHz). After doing this experiment (MNG,ENG,DNG) the receive power of LR (MNG), Wire (ENG) and the LR and Wire (DNG magnitude and phase) are plotted in the Fig.3(a) and 3(b).

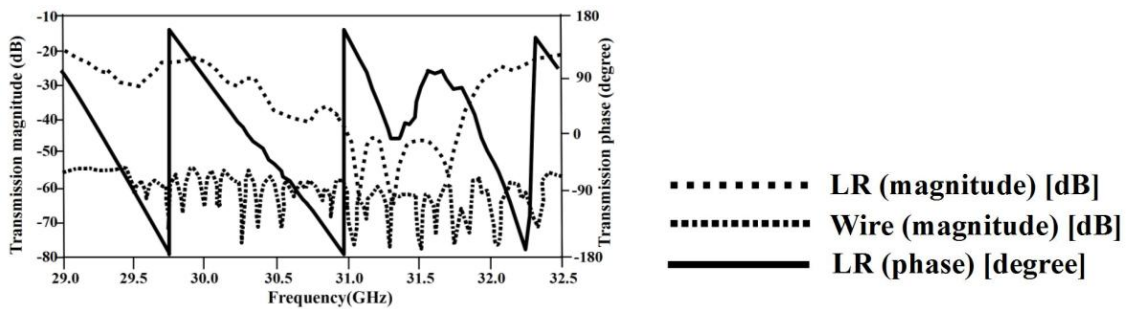

(a)

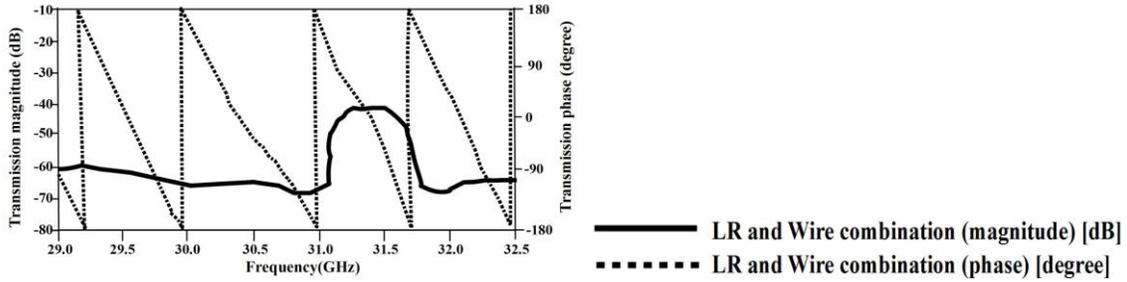

**Fig.3 (a)The transmission experiment of LR (magnitude),Wire (magnitude)**

(b)
3. (b) LR-Wire combination-DNG (magnitude and phase).

While performing this experiment we observed that in the LR experiment there, is a stop band observed in the range of (about) 31GHz to 32GHz. Similarly for characterization of WA, the (transmission) experiment shows an entire stop band in the entire Ka band. Actually, the WA structure gives stop band below 42 GHz, in our case. The combination of LR and Wire gave DNG medium there is a band pass is formed in the range of (about) 31-32 GHz; (where the stop band is observed in LR-characterization). In the Fig.3 (b) LR-Wire combination which realizes DNG gives the phase reversal in the band of (about) 31-32 GHz, the range where the NRM phenomena are occurring. Fig.4 gives that how this phase reversal phenomena occur in the NRI medium and it is comparable with ordinary material, indicates that travelling wave phases are opposite inside NRM, vis-à-vis normal media, and that is phase velocity inside NRM is opposite to that in air. The experimental results of LR, Wire and DNM material are given in Table-3. Our experimental results are closely matched with reference [18] wherein, the author gives a new structure and formulation for MTM experiment.

**Table: -3**

| *Inclusion Structures* | *LR Array (MNG)* | *WA (ENG)* | *LR-WA Combination (DNM)* |
|---|---|---|---|
| Frequency Band (GHz) | 30.85-31.9 GHz (Stop Band) | 0-42 GHz (Stop Band) | 31.1-32 GHz (Pass Band) |
| Magnitude value (dB) | 20 dB (below the upper) | 55dB (below the upper) | 20 dB ( below ground floor) |
| Phase change (degree) | $0^o$ phase change | $90^0$ phase change (approximately) | $0^o$ phase change |

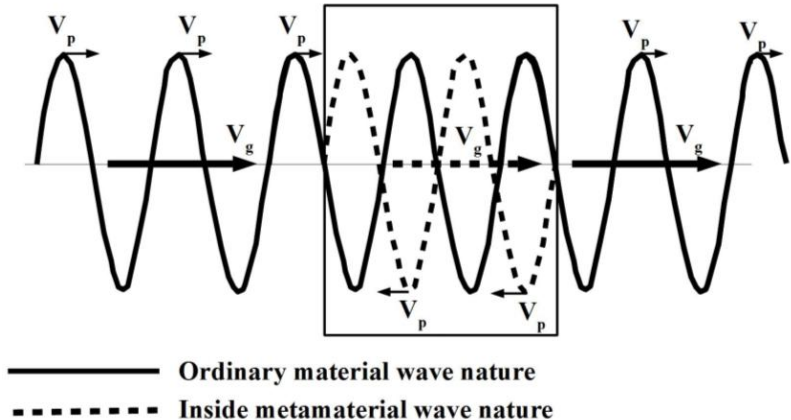

**Fig.4 Schematic diagram of wave behavior inside an ordinary and metamaterial substance.**

When a electromagnetic signal enters into the ordinary material the phase velocity (VP) and group velocity (VG) are in phase so the equation is formed $V_p \cdot V_g = c^2$ [7] Inside the metamaterial the phase and group velocity are $180^o$ apart from each other, and we have $V_p \cdot V_g = -c^2$, this is true for the $n = -1$ case, depicted in fig-4. The polarization of the antenna and direction of propagation from this transmission experiment are given in Fig.2.

### III. Negative Refraction Index (NRI) Experiment

The NRI experiment is performed in various ways [19-21] but researchers there are doing it by cutting their DNG material to either wedge shape or by making prism. For performing this NRI experiment researchers used Horn lens antenna which is very costly and also bulky in Ka band. The novelty of our experiment is that we can do the same with ordinary horn antenna and get an indirect method of finding the NRI of the DNG material. We used Far field fronts instead of using horn lens antenna. Horn antenna in the near field has its radiations as spherical in nature and in far field it's a plane wave . We have placed our DNG material (Prism shape) in a wooden box (inside is covered up with absorber and outer part cover up with the absorber as in Fig.6). Therefore at the far field, from the horn antennas, very small portion is incident on the NRI material hence, effectively the wave is formed is a plane wave. This arrangement gives two advantages as follows:

1. The horn lens antennas are big and bulky in nature of Ka band its size is approximately 15 inches and its front diameter its 10 inches, and you have to place at some distance for producing of plane beam purpose. Though its size is an issue but it has some advantages, using this we can achieve high gain and low side lobes.

    The criteria for Far field is to place the receiving antenna very very greater than $2x\lambda_0$. Despite of this in our experiment with an ordinary horn which have less gain compare to horn lens antenna and we placed it in the Far field region (for achieving the plane wave ) we take the distance between the transmitter antenna and the NRI material distance is $41x\lambda_0$ where the value of $\lambda_0 = 0.97$ cm (central frequency 31GHz). In our experiment we covered the NRM material with microwave absorber so very less power is incident on the NRI material (the horn has low gain) its power level is very low, but we are measuring the power level from its ground level or we can say that we are measuring the power level difference with respect to ground level, so we say that it is invariant of its gain. So the benefit of our experiment is by using a low cost antenna and some microwave absorber we can do our NRI experiment. (See Figure 5a)

2. The transmitted wave is bound to an incident on the DNM because it is placed in the far field and if a small portion of a far field EM wave is incident on the DNM then it will behave like a plane wave. The experimental setup and the wooden box are covered by the microwave absorber so there is no chance for any external wave interfere
.

3. Far field region the waves are plane in nature so there electric and magnetic field and the propagation of directions are perpendicular to each other. The choice of wooden box is that it has no effect on electric as well as in a magnetic field. For better accuracy we cover the inside with the absorber and as well as in the outside, we placed the DNM inside this wooden box that will serve the better accuracy on the experiment, Fig.5.

For the upper discussion is the merit of our experiment. In the Fig.5. gives the schematic representation of our experimental and the conventional horn antennas representation.

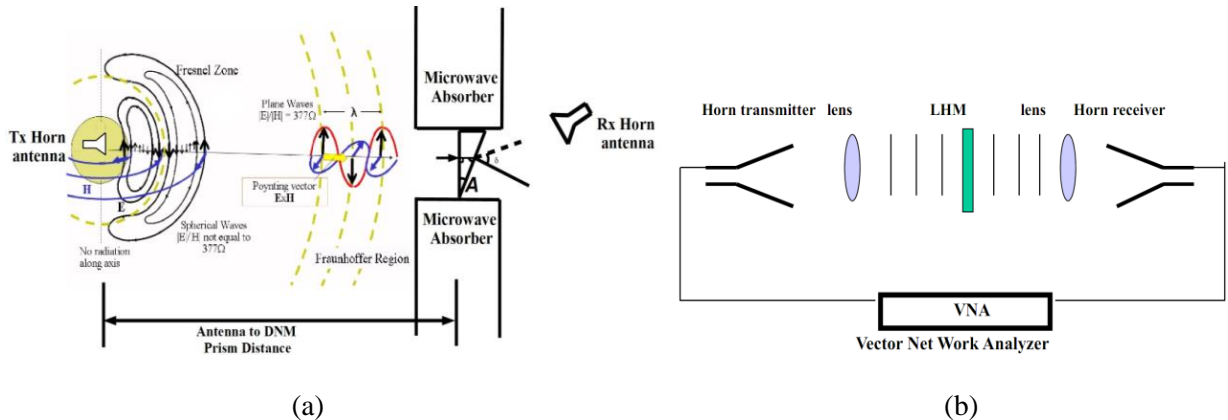

(a) (b)

**Fig.5. (a) Schematic representation of the proposed experiment. (b) Horn Lens antenna Experiment.**

To perform this negative refractive index (NRI) material experiment first we made our (LR-Wire combination-DNM) prism $45^0$ and we observed that there are two regions where we get the same power (+ve and –the refraction region). We therefore concluded that for $45^0$ prism is actually a 'power splitter' that's why it's dividing input power into two parts(+ve and –ve refraction region). After this experimental ambiguity (and its realization) we made our prism for $15^0$ angles performed the experiment [10]. The $15^0$ prisms are given in Fig.6 (a) and the experimental arrangement enclosed by microwave absorber, to avoid errors due to external electromagnetic interference. This experiment is given in the Fig.6 (b), the internal arrangement of wooden box for placing the prism (DNM) is given in Fig.6 (c).

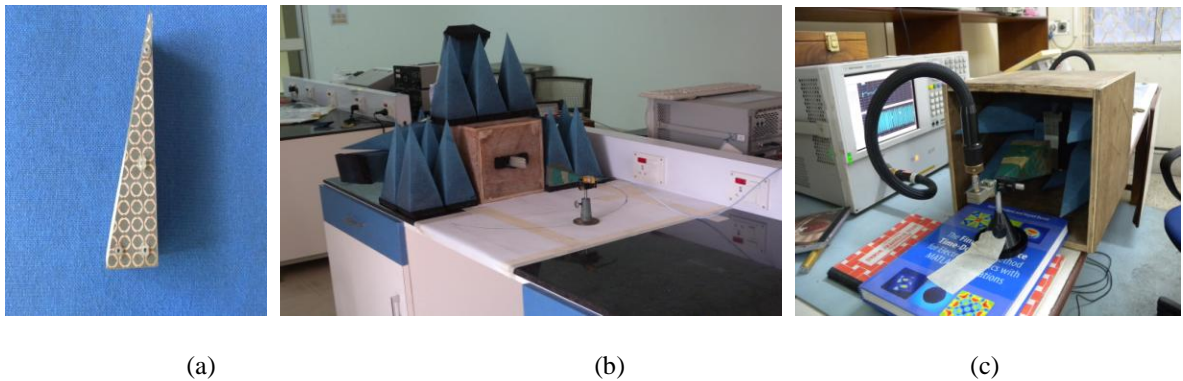

(a) (b) (c)

**Fig.6 (a) $15^0$ Prism of NRM experiment. (b) Experimental arrangement for NRM experiment. (c) Wooden box internal arrangements.**

Usually for finding the refractive index experiments researchers are done by changing angle of the receiver, with a constant frequency of the source transmitter. The method is to be found the angle where the receiver gets maximum output power. In our experiment we do the reverse process, we change the receiving antenna angle to the entire frequency range (ka band 26.5 GHz-40 GHz), and we want to find the receiver angle with a sudden frequency where the NRI material can pass its maximum energy with some sudden negative angle. So at that frequency (highest NRI frequency) and that angle we get an NRI pass band.

In our experiment we observed that there is an angle ($-29.28^0$) where the maximum power is obtained at this frequency range (31-32 GHz). Fig.6 gives the frequency 31.45 GHz where the maximum level of power flow is observed. The difference between the peak and the ground level is approximately 20 dB. This method is not described in any article for measuring the refractive index of the NRI material. At any other angle (other than $-29.28^0$) we do not observe this peak transmittance. Fig.6. gives it frequency verses power with some centering frequency. The calculation by using a prism (NRI) is given in below.

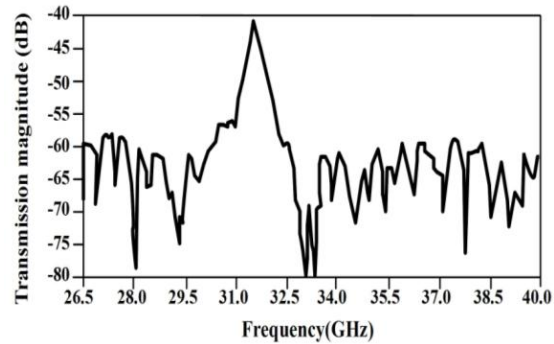

**Fig.6. Transmission experiment for observing the NRM angle.**

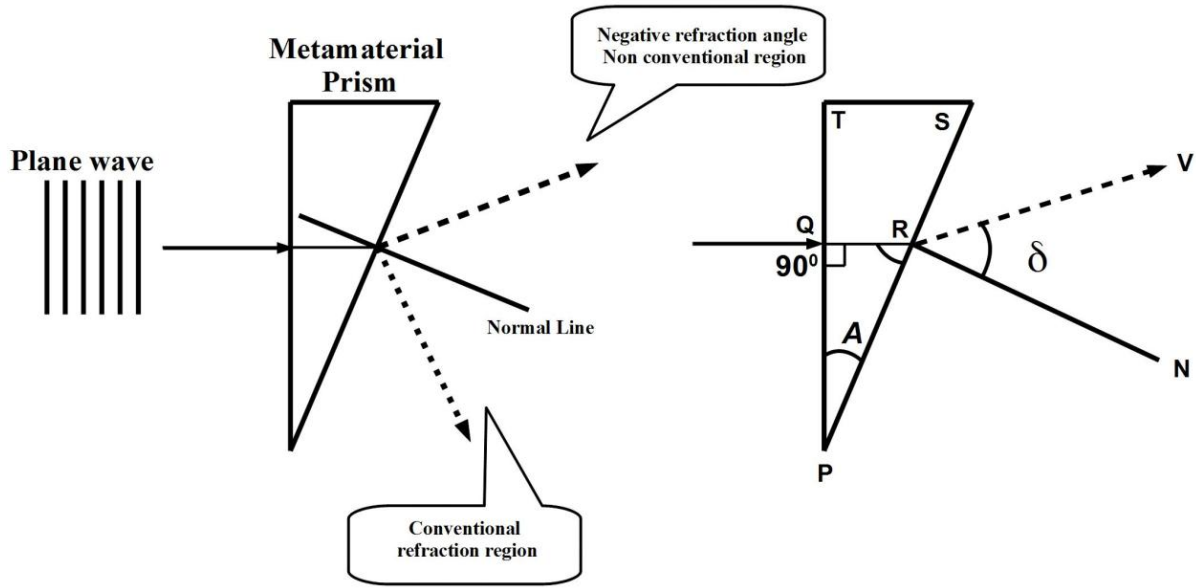

**Fig.7. Basic Prism Geometry for NRI experiment.**

The prism refractive index calculation and its deviations are given by reference [22]. This reference is concerned with the positive index material not for negative indexing material. For calculation of NRI are given by many researcher Ekmel et.al. [23] one of them. In our experimental characterization we take our own approach for calculating the NRI of DNM (LR and WA). The prism following details are as:-

$\angle QRA$ =incident angle and $\angle VRN$ =deviation angle.

Now the $n_{eff}$ is the material effective refractive index and $n_{air}$ is the air refractive index which is one. In this prism experiment we assume that first the plane wave are inserted into the metamaterial where it will show the reversed *snell law* and after that in the air medium so from the *Snell's law* we can write as;

$$n_{eff} \sin(\angle incident) = n_{air} \sin(\angle refracted) \qquad (2)$$

Where $n_{air}=1$.

$$n_{eff} = \frac{\sin(\angle refracted)}{\sin(\angle incident)} \qquad (3)$$

*incident angle*= $\angle QRA$ =15° and *refractive angle*= $\angle VRN$ = -29.28° so the negative refractive index $n_{eff}$ is -1.89.

In equation (1) and the wire analytical expression using MATLAB simulation we can get that in unit cell analytical simulation the value of effective NRI is -1.88 which is very close to our experimental results. The Fig.8. give the analytical simulated result of LR and wire at the entire Ka band.

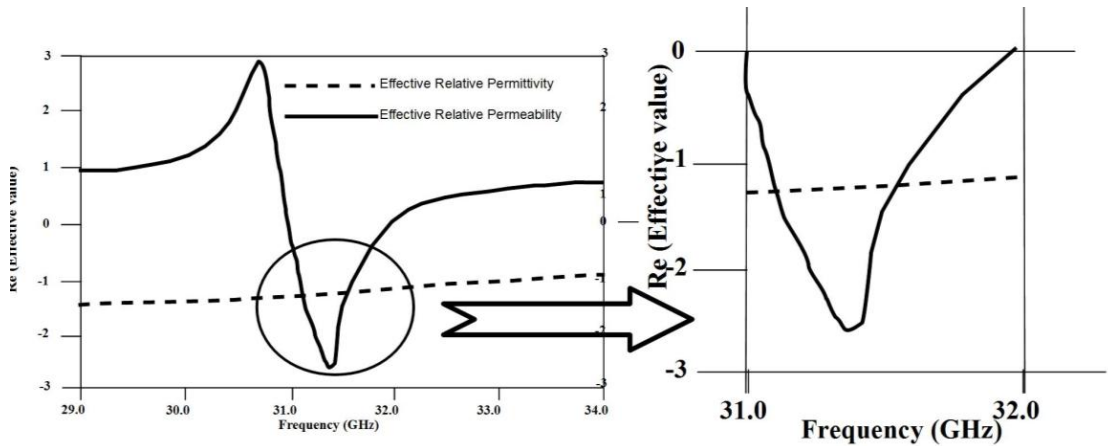

Fig.8. Analytical results of LR and Wire using MATLAB.

[ From the upper figure it is found that 31.45GHz the effective relative permeability($\mu_{reff}$) value is -2.71 and effective relative permeability is ($\varepsilon_{reff}$) is -1.30 so from the refractive index equation is $n_{reff} = \sqrt{\varepsilon_{eff} \mu_{eff}}$ so the value of NRI for this DNM is:- $n_{reff} = \sqrt{-\varepsilon_{eff}} \sqrt{-\mu_{eff}} = -\sqrt{\varepsilon_{eff} \mu_{eff}} = -1.88$ ]

The LR has basic differences with SRR (Split Ring Resonator) that is in LR there is two gap in a single ring. From our analytical study we learn that if the gap number is increase in a single ring then its effective relative permeability region will increase on the other hand we say 180° rotation of symmetry in plane element, as a consequence of this symmetry, cross polarization effects are eliminated, that's why we chose LR from our MNG material [24].

### IV. CONCLUSION

In this paper firstly we reported the transmission experiment for characterization of NRI material and from there we conclude that approximately for 1GHz of the bands where the both LR and WA are permeability and permittivity both are negative or vis a vis we say that this frequency range its refractive index value become negative. Though the experimental characterization is the main goal of this paper that's why we cannot elaborate the analytical expression of LR and WA. The fabrication of LR and WA are done by using the basic analytical expression which is reported by many authors. In the last portion we have already discussed that due to cross polarization effect the LR gives more effective relative permeability region compare to any other conventional MNG material. In our experiment we observed that approximately 1GHz of NRI band is formed which is matching with our analytical results(Fig.8). In WA its pass band is up to 42GHz, so there is no problem of using it, with our LR structure.

In the experiment we have used the ordinary horn antenna that produced a plane wave in the far field region. The main intention of the experiment is to find the maximum NRI of a DNM. The researcher used a Horn lens antenna to find the entire band NRI response. The horn lens antennas have high gain and low side lobes but its size and its cost not feasible. In this experiment we do it in the far field region and we make a small part of wave ,incident on the sample .In our experimental point of view we want the specific frequency where the NRI has maximum value. Our experiment we observed that around -40dB is the peak point at 31.43GHz and its ground floor is at -60dB. So we can conclude that the angle of $-29.28^0$ the EM wave pass and it will come with the receiving horn antenna and after that using the simple prism geometry we can find the value of NRI of the DNM. If we had used the Horn lens antenna for this experiment then we would observe that the power level increase because the lens antennas basics are to converge all the side lobe to the main lobe. The using of microwave absorbing the unwanted wave of both inside and outside both are eliminated. The wedge experiment is similar to the prism experiment and using this wedge experiment we can also find the NRI value of DNM.

In the experiment the LR and WA in the vertical lattice constant we inserted U-Foam material whose dielectric constant is same as air, gives mechanical stability. Though it is very good because its operation frequency is very large but its drawback is it is a very lossy material so that will also cause to increase more losses and also using two dielectric for separated for LR and WA that will also cause the more losses. The one solution is that using a single dielectric if we imprinted both the LR and WA then half amount of loss will decrease so probably more output power can achieve in the experiment.

**References**:

1. V.G.Veselago, "The electrodynamics of substances with simultaneously negative values of $\varepsilon$ and $\mu$", Usp.Fiz.Nauk 92, 517-526, July 1964.
2. Sougata Chaterjee, Arijit Mazumder ,Amitesh Kumar, Paulami Sarkar,Ananta Lal Das, Shantanu Das, and Subal Kar, " Particle Energy Momentum Transportation for Negative Refractive Index Material(NRM) Anamolous Concept", ISAP-IEEE-2011 Conference.
3. Shantanu Das, Review on Composite Negative Refractive Indexed Material with Left Handed Maxwell Systems, Fundamental Journal of Modern Physics, Vol.3, Issue-1,pp 13-89, [2012].
4. Shantanu Das, Electromagnetic Momentum and Energy inside Negative Refractive Indexed Material: A new look at concept of photon -Invited Talk Physical Science Section ISC, Indian Science Congress 99th Session Bhubaneswar, [2012].
5. Shantanu Das, "Quantized Energy Momentum and Wave for an Electromagnetic pulse-A single photon inside Negative Refractive Indexed Media", Journal of Modern Physics 2011 Vol. 2 No.12, pp 1507-1522, [2011]
6. Shantanu Das, Sougata Chaterjee, Amitesh Kumar, Arijit Mazumder , "A New Mechanics of Corpuscular-Wave Transport of Momentum and Energy inside Negative Indexed Material", , Fundamental Research and Development International "Fundamental Journal of Modern Physics"; Vol.1, Issue2, pp.223-246, [2011]
7. Shantanu Das, Sougata Chaterjee, Amitesh Kumar, Paulami Sarkar, Arijit Majumder, Ananta Lal Das, Subal Kar; "A new look at the nature of linear momentum & energy of electromagnetic radiation inside negative refractive indexed medium", , Physica Scripta 84 (2011) 035707 8pp. August 2011, [2011]

8. J.B. Pendry, A.J.Holden, and W.J.Stewart, "Extremely low frequency plasmons in metallic mesostructures", Phy. Rev.Lett, Vol.76.No.25,4773-4776, June 1996.
9. J.B.Pendry, A.J.Holden, D.J.Robbins, and W.J.Stewart, "Magnetism from Conductors and Enhanced Nonlinear Phenomena", IEEE Trans MTT,Vol.47,pp.2075-84,Nov.1999.
10. Shelby, R. A., D. R. Smith, and S. Schultz, "Experimental verification of a negative index of refraction," Science, Vol. 292, No. 6, 77–79, 2001.
11. L. Ran, J. Huangfu, H. Chen, X. Zhang, K. Cheng,T. M. Grzegorczyk and J. A. Kong "Experimental study on several left-handed metamaterials" PIER 51, 249–279, 2005.
12. J.D. Baena, R. Marqués1, J. Martel, and F. Medina,"Experimental results on metamaterial simulation using SRR-loaded waveguides",Proc.IEEE-AP/S Int.Symp.on Antenna Propagation,pp-106-109,2003.
13. Andrew A. Houck, Jeffrey B. Brock, and Isaac L. Chuang, "Experimental Observations of a Left-Handed Material That Obeys Snell's Law" PH YSICA L R EV I EW L ET T ERS, VOLUME 90, NUMBER 13.
14. C. D. Moss, T. M. Grzegorczyk, Y. Zhang, and J. A. Kong "NUMERICAL STUDIES OF LEFT HANDED METAMATERIALS" PIER 35, 315–334, 2002.
15. Irfan Bulu, Humeyra Caglayan, and Ekmel Ozbay "Experimental demonstration of labyrinth-based left-handed metamaterials" Optics Express, Vol. 13, Issue 25, pp. 10238-10247 (2005).
16. T. Roy, D. Banerjee, S. Kar," Studies on Multiple-Inclusion Magnetic Structures Useful for Millimeter-Wave Left-Handed Metamaterial Application " IETE Journal of Research,Vol-55,Issue-2,Mar-Apr 2009.
17. D.R.Smith, S.Schultz, P.Markos, and M.Soukoulis, "Determine of effective permittivity and permeability of metmaterial from reflection and transmission coefficients", Physical Review B.Vol.65,195104,April 2002.
18. Richard W. Ziolkowski ,"Design fabrication and testing of double negative metamaterial", IEEE -Antenna and Propagation, July-2003,pp-1516-1529.
19. Andrew A. Houck, Jeffrey B. Brock, and Isaac L. Chuang, "Experimental Observations of a Left-Handed Material That Obeys Snell's Law" PH YSICA L R EV I EW L ET T ERS, VOLUME 90, NUMBER 13.
20. C. D. Moss, T. M. Grzegorczyk, Y. Zhang, and J. A. Kong "NUMERICAL STUDIES OF LEFT HANDED METAMATERIALS" PIER 35, 315–334, 2002.
21. Jiangtao Huangfu, Lixin Ran, Hongsheng Chen, Xian-min Zhang, Kangsheng Chen," Experimental confirmation of negative refractive index of a metamaterial composed of Ω-like metallic patterns", Appl. Phys. Lett. 84, 1537 2004); doi: 10.1063/1.1655673.
22. HonacB WrNcuntl, Yale UniaersityN, ew Haaen,C onm "ALIGNMENT CHART FOR CALCULATION OF REFRACTIVEINDEX FROM THE DEVIATION OF LIGHT BY A PRISM".
23. Ekmel Ozbay, Kaan Guven, and Koray Aydin,"Metamaterials with negative permeability and negative refractive index: experiments and simulations", J. Opt. A: Pure Appl. Opt. **9** (2007) S301–S307.
24. Sougata Chatterjee, Arijit Majumder, Amitesh Kumar, Paulomi Sarkar, Ananta Lal Das, Shantanu Das, Subal Kar "Analytical and Simulation Study on Different Magnetic Inclusion Structures" ISAP-IEEE-2011 Conference.